\documentclass[aps,prb,reprint,groupedaddress]{revtex4-2} 
\pdfoutput=1
\usepackage{siunitx}
\usepackage{graphicx}
\usepackage[
outdir=./fig
]{epstopdf}
\usepackage{amsmath, amssymb, nicefrac}
\usepackage{booktabs}

\newcommand{\lwr}[1]{\textsubscript{\protect\raisebox{-1pt}{#1}}}
\newcommand{\upr}[1]{\textsuperscript{\protect\raisebox{1pt}{#1}}}
\newcommand{\mlwr}[1]{_{\mathrm{#1}}}			
\newcommand{\mupr}[1]{^{\mathrm{#1}}}
\newcommand{\uplwr}[2]{\rlap{\upr{#1}}\lwr{#2}}

\begin{document}
	
	\title[Defect conversion in SiC]{Conversion pathways of primary defects by annealing in proton-irradiated $n$-type 4H-SiC}
	\author{Robert Karsthof}
	\affiliation{Centre for Materials Science and Nanotechnology, Universitetet i Oslo, Gaustadalléen 23A, 0373 Oslo, Norway}
	\email{r.m.karsthof@smn.uio.no}
	
	\author{Marianne Etzelm\"{u}ller Bathen}
	\affiliation{Centre for Materials Science and Nanotechnology, Universitetet i Oslo, Gaustadalléen 23A, 0373 Oslo, Norway}

	\author{Augustinas Galeckas}
	\affiliation{Centre for Materials Science and Nanotechnology, Universitetet i Oslo, Gaustadalléen 23A, 0373 Oslo, Norway}

	\author{Lasse Vines}
	\affiliation{Centre for Materials Science and Nanotechnology, Universitetet i Oslo, Gaustadalléen 23A, 0373 Oslo, Norway}

	\date{\today}
	
\begin{abstract}

	The development of defect populations after proton irradiation of $n$-type 4H-SiC and subsequent annealing experiments is studied by means of deep level transient (DLTS) and photoluminescence (PL) spectroscopy. A comprehensive model is suggested describing the evolution and interconversion of irradiation-induced point defects during annealing below \SI{1000}{\degreeCelsius}. The model proposes the EH\lwr{4} and EH\lwr{5} traps frequently found by DLTS to originate from the (+/0) charge transition level belonging to different configurations of the carbon antisite-carbon vacancy (CAV) complex. Furthermore, we show that the transformation channel between the silicon vacancy (V\lwr{Si}) and CAV is effectively blocked under $n$-type conditions, but becomes available in samples where the Fermi level has moved towards the center of the band gap due to irradiation-induced donor compensation. The annealing of V\lwr{Si} and the carbon vacancy (V\lwr{C}) is shown to be dominated by recombination with residual self-interstitials at temperatures of up to \SI{400}{\degreeCelsius}. Going to higher temperatures, a decay of the CAV pair density is reported which is closely correlated to a renewed increase of V\lwr{C} concentration. A conceivable explanation for this process is the dissociation of the CAV pair into separate carbon anitisites and V\lwr{C} defects. Lastly, the presented data supports the claim that the removal of free carriers in irradiated SiC is due to introduced compensating defects and not passivation of shallow nitrogen donors.

\end{abstract}

\maketitle

\section{Introduction}

Silicon carbide (SiC) possesses a variety of point defects and defect complexes that are promising for quantum communication and quantum computing applications, due to both the emission of single photons upon optical excitation and the existence of high-spin states for electrons being trapped at these defect sites. Among the most studied polytypes (3C, 4H, 6H), 4H-SiC possesses the largest band gap of \SI{3.23}{\electronvolt} at room temperature. It can be grown with acceptably low residual impurity concentrations such that intrinsic defects play a dominant and controllable role in material properties. Moreover, many of its intrinsic defects that can be introduced by electron or ion irradiation, or implanation, have been shown to be excellent candidates for quantum building blocks \cite{Lohrmann.2017}.\\
Herein, we mainly consider six intrinsic defects in 4H-SiC that are electrically and/ or optically detectable: the carbon and silicon vacancies (V\lwr{C}, V\lwr{Si}) and self-interstitials (C\lwr{i}, Si\lwr{i}), the carbon antisite-carbon vacancy pair (C\lwr{Si}V\lwr{C}), and the divacancy (V\lwr{C}V\lwr{Si}). The V\lwr{C} has been shown to have formation energies below \SI{5}{\electronvolt} \cite{Szasz.2015,Kobayashi.2019}, and therefore is expected to be present in as-grown SiC, in accordance with experimental observations \cite{Son.2007,Zippelius.2012}. In 4H-SiC, V\lwr{C} can occur in two crystallographically inequivalent lattice sites labeled $h$ and $k$ for pseudo-hexagonal and pseudo-cubic, respectively, and the formation energies of these two defects can differ by several \SI{100}{\milli\electronvolt}, depending on the charge state \cite{Szasz.2015}. V\lwr{C} also exhibits strong negative-$U$ effects and a pronounced Jahn-Teller distortion of the singly-negative charge state \cite{Trinh.2013,Coutinho.2017}, leading to the direct transition from the neutral to doubly-negative (0/2$-$) acceptor state at around \SI{0.7}{\electronvolt} below the conduction band edge $E\mlwr{C}$ \cite{Son.2012}. V\lwr{C} seems to be an efficient center for non-radiative recombination and is therefore considered a lifetime-limiting defect \cite{Ayedh.2017} which has detrimental implications for bipolar devices.\\
The formation energy of the silicon monovacancy V\lwr{Si} spans over a large range from around \SI{7.5}{\electronvolt} under $p$-type and intrinsic conditions to as low as \SI{4.5}{\electronvolt} in $n$-type SiC \cite{Szasz.2015, Kobayashi.2019}. V\lwr{Si} is mainly acceptor-like, possessing charge states between neutral and threefold negative, although a shallow donor-like ${(+/0)}$ transition close to the valence band is possible \cite{Szasz.2015, Kobayashi.2019}. Analogously to V\lwr{C}, two configurations of V\lwr{Si}($h,k$) exist. Charge transition levels (CTLs) for the V\lwr{Si} occur at around $E\lwr{C}-\SI{0.7}{\electronvolt}$ for the ${(-/2-)}$ and $E\lwr{C}-\SI{0.4}{\electronvolt}$ for the {(2$-$/3$-$)} levels \cite{Szasz.2015,Bathen.2019}. V\lwr{Si}, in contrast to V\lwr{C}, does not exhibit negative-$U$ character. The singly-negative charge state of V\lwr{Si} is optically addressable with a spin of $S = \nicefrac{3}{2}$ that has been shown to have a long spin coherence time \cite{Nagy.2019}. This makes it an ideal candidate for both solid-state qubits as well as single-photon sources, given its favorable emission wavelengths in the near-infrared spectral range (V lines). V\lwr{Si} defects can be controllably introduced into SiC by proton irratiation \cite{Kraus.2017} and its charge state can be controlled using, e.g., Schottky barrier diodes \cite{Bathen.2019}.\\
The carbon antisite-carbon vacancy pair (CAV pair) is considered the counterpart to V\lwr{Si} \cite{Steeds.2009}, because it can be formed by a single carbon hop into a neighboring V\lwr{Si}. In $p$-type material, the formation energies of the CAV pair are significantly lower than that of the V\lwr{Si} \cite{Szasz.2015}. At around \SI{1}{\electronvolt} below the conduction band edge $E\mlwr{C}$, the situation is reversed, with V\lwr{Si} becoming thermodynamically slightly more favorable. Because the CAV pair involves one site from the C and another from the Si sublattice, four inequivalent configurations of the CAV pair can be realized, termed $hh$, $hk$, $kh$ and $kk$, adapting the notation for the monovacancies and referring to the Si site with the first and the C site with the second symbol, respectively. The CAV pair possesses three CTLs within the 4H-SiC band gap, two of which are donor-like ((2+/+) and (+/0) transitions at roughly mid-gap and \num{1.0}-\SI{1.1}{\electronvolt} below $E\mlwr{C}$, respectively) and one is acceptor-like (${(0/-)}$ CTL, around \SI{0.5}{\electronvolt} below $E\mlwr{C}$). The CAV pair in its singly-positive charge state is known to be an ultra-bright single-photon source, emitting in the visible spectral range \cite{Castelletto.2013} (AB lines).\\
It shall be noted that the silicon antisite-vacancy pair (Si\lwr{C}V\lwr{Si}) is excluded from the list of relevant defect complexes here because it has been shown to be unstable with regard to V\lwr{C} \cite{Bockstedte.2003}.\\
Formation of the divacancy V\lwr{Si}V\lwr{C} (VV) \cite{Son.2006,Son.2007} requires not only abundance of single Si and C vacancies, but also their adjacency which can be achieved at elevated temperatures via defect diffusion. Therefore, although its formation energy has been calculated to be in the same range as for V\lwr{Si} and the CAV pair \cite{Wang.2013}, the equilibrium concentration of VV defects depends on the extent to which the single vacancies (mostly V\lwr{Si}, due to their lower migration barriers as compared to V\lwr{C} \cite{Defo.2018}) are allowed to diffuse, for example through post-annealing or during high-temperature growth or irradiation. Thermodynamic CTLs for the VV defect have been calculated in the literature to be in the range of \num{0.7}-\SI{0.9}{\electronvolt} and \num{1.2}-\SI{1.4}{\electronvolt} below $E\mlwr{C}$ for the ${(-/2-)}$ and ${(0/-)}$ CTLs, respectively \cite{Gordon.2015,Magnusson.2018,Csore.2019}. Analogously to the CAV pair, being a complex comprising two lattice sites, there are four non-degenerate configurations of the VV as well, yielding an energetic spread of each of the CTLs. The VV defect in the neutral charge state is also a known single-photon source in 4H-SiC, emitting in the infrared range around \SI{1.2}{\electronvolt} \cite{Magnusson.2018,Anderson.2019}.\\
Even though many routes have been explored on how to introduce and control certain defects by irradiation and subsequent annealing, a complete picture of the conversion pathways through which defects can be interconverted, and under which conditions this occurs, is still missing. Such a general model can be expected to be indispensable when developing post-irradiation annealing strategies to achieve a specific population of defects. This paper aims at taking a step towards such a model by studying epilayers that are proton-irradiated to a wide range of fluences, and thereafter annealed in isochronal steps to increasingly higher temperatures. In each step, the defect population is monitored by means of deep level transient spectroscopy (DLTS) and photoluminescence (PL) spectroscopy in order to reveal defect interconversion and out-annealing processes.

\section{Experimental methods}
	
The investigated samples consisted of \SI{10}{\micro\meter} thick, nitrogen-doped, (0001) oriented epitaxial 4H SiC layers, purchased from Cree Inc. The doping concentration was around $N\mlwr{D} \approx \SI{1e15}{\per\cubic\centi\meter}$, as determined by capacitance-voltage ($C$-$V$) measurements. The $n$-doping of the SiC substrate amounted to about \SI{8e18}{\per\cubic\centi\meter}. Formation of intrinsic defects was achieved by room temperature irradiation with \SI{1.8}{\mega\electronvolt} protons, having a projected range of \SI{27}{\micro\meter}, well outside the epilayer, as simulated based on collision Monte Carlo models implemented in the SRIM package \cite{Ziegler.2010}. In order to suppress channeling effects, the incident beam was tilted by \SI{8}{\degree} with respect to the surface normal. Irradiation fluences were chosen between \SI{1e11}{\per\square\centi\meter} and \SI{6e13}{\per\square\centi\meter}. This lead to two subsets of samples: one in the fluence range of up to \SI{4e12}{\per\square\centi\meter}, with trap concentrations being low enough ($< 0.2 N\mlwr{D}$) to perform DLTS measurements, and one being subjected to higher fluences that exhibited defect concentrations above the lower detectivity threshold for PL measurements. Moreover, the higher-fluence set showed full donor compensation and was therefore not suitable for DLTS initially. \\
After irradiation the samples were thermally annealed at \SI{300}{\degreeCelsius} in a tube furnace in flowing N\lwr{2} (\SI{30}{\cubic\centi\meter\per\minute}) to alleviate the influence of irradiation-induced unstable defects. In order to perform defect spectroscopy measurements, Schottky barrier diodes (SBDs) were fabricated on the epilayer surface, using electron beam-evaporated nickel patterned by deposition through a shadow mask. Before SBD fabrication, the samples underwent RCA cleaning to remove contaminations, including any SiO\lwr{2} surface layers. The SBDs were of circular cross-sections and had an area of \SI{7.85e-3}{\square\centi\meter}, with a Ni thickness of \SI{150}{\nano\meter}. The samples intended for PL measurements were thoroughly cleaned in an ultrasonic bath in acetone and isopropyl alcohol for 5 minutes each, but were left untreated otherwise.\\
In order to study the influence of thermal treatment on defect conversion, isothermal annealing at increasingly higher temperatures and subsequent DLTS characterization was performed. This included the removal of SBDs before each annealing step to avoid in-diffusion and alloying of Ni into the SiC surface. For that purpose, the samples were placed in a 5:1:1 mixure of DI-H\lwr{2}O:HCl:H\lwr{2}O\lwr{2} at a temperature of \SI{80}{\degreeCelsius} for ten minutes, which leaves the SiC free of metal residues. After annealing, the samples were quenched to room temperature using a cold plate, and the SBDs were re-applied. The annealing steps took place at \SI{400}{\degreeCelsius}, \SI{600}{\degreeCelsius}, \SI{800}{\degreeCelsius} and \SI{1000}{\degreeCelsius}, otherwise analogously to the conditions described for the first step above.\\
DLTS measurements were carried out using a high-temperature setup operating between \SI{77}{\kelvin} and \SI{700}{\kelvin}. The measurement frequency was \SI{1}{\mega\hertz}, the pulse length was \SI{20}{\milli\second} and the pulse height \SI{10}{\volt} at a reverse bias of \SI{-10}{\volt}. Six rate windows between \SI{20}{\milli\second} and \SI{640}{\milli\second} were chosen for evaluation of the transients, using a standard lock-in correlation function. The concentrations of the different levels found in the measured spectra were extracted by a numerical simulation of the transients and the resulting spectra, based on assumed (variable) values for trap energy, capture cross section and density of the traps. The lambda correction was included for increased accuracy of the thusly determined trap concentrations. \\
Photoluminescence measurements were carried out using a closed-cycle He refrigerator system (Janis, CCS450) and a \SI{405}{\nano\meter} wavelength cw-laser of \SI{75}{\milli\watt} power as excitation source. The focused laser beam, impinging on the sample surface at a \SI{27}{\degree} angle, yielded an excitation intensity of $\le \SI{1}{\kilo\watt\per\square\centi\meter}$ and polarization perpendicular to the optic $c$-axis of 4H-SiC. The PL signal was collected in a back-scattering geometry by a microscope objective (Mitutoyo, LWD 10X), spectrally filtered (long-pass LP \SI{550}{\nano\meter} filter) and analyzed by an imaging spectrometer (Horiba, iHR320) coupled to an EMCCD camera (Andor, iXon Ultra 888) with a spectral resolution below \SI{0.2}{\nano\meter}. A near-confocal configuration of the detection, ensured by a high numerical aperture objective and narrow slit of the imaging spectrometer, allowed for maximized collection of the PL signal from the uppermost 3--\SI{4}{\micro\meter} of the epilayer.

\section{Results and Discussion}

\subsection{Assessment of initial defect populations}

\begin{figure}
	\centering
	\includegraphics[width=\columnwidth]{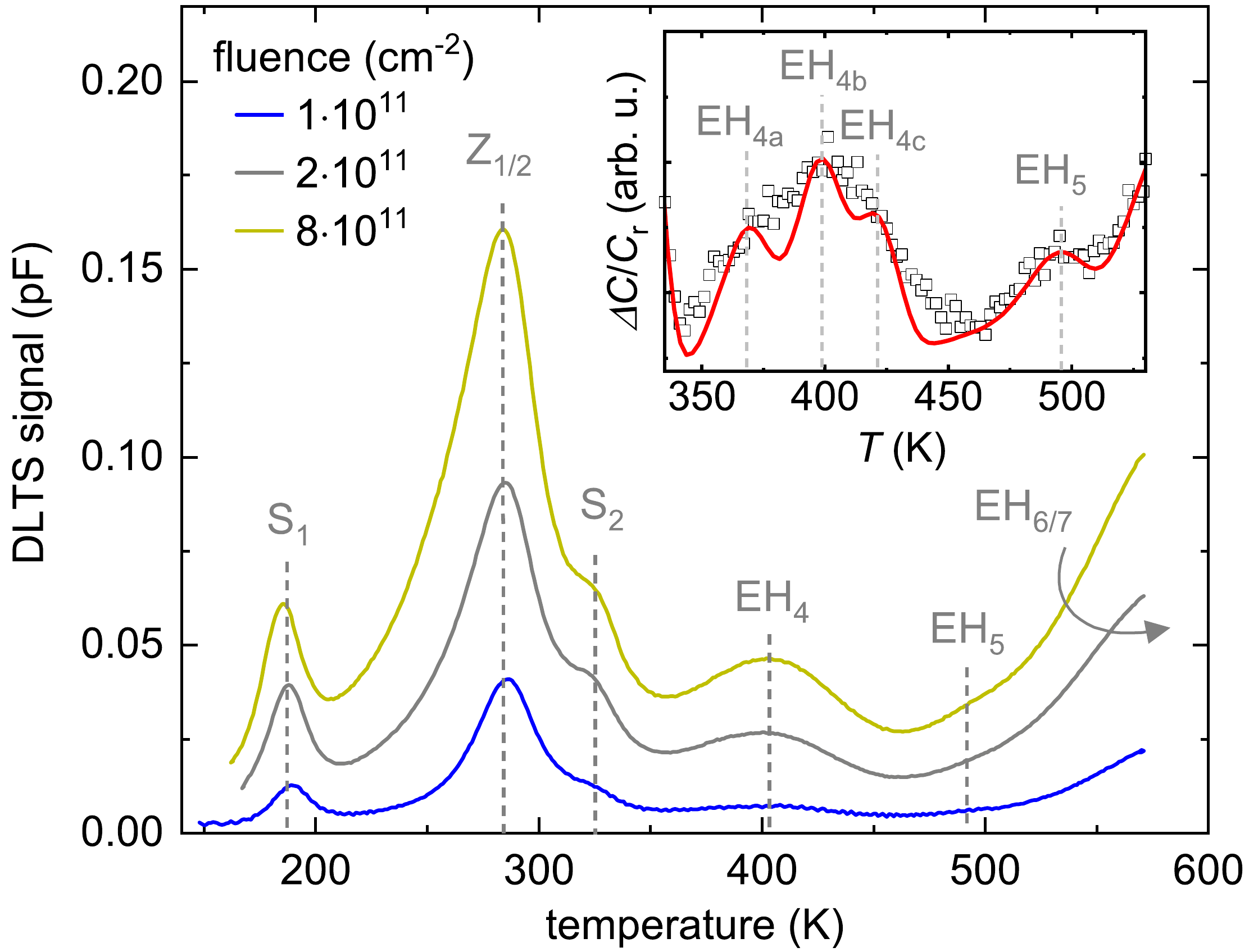}
	\caption{DLTS spectra of 4H-SiC proton-irradiated to three different fluences after an initial annealing step to \SI{300}{\degreeCelsius}, with dominant defect levels indicated. The corresponding rate window was \SI{1.56}{\per\second}. Inset: high-resultion DLTS spectrum (fluence \SI{8e11}{\per\square\centi\meter}) showing the temperature region of the EH\lwr{4} and EH\lwr{5} traps, revealing three contributions to EH\lwr{4}.}
	\label{fig:DLTS_P1_P3_P4_300C}
\end{figure}

Fig.~\ref{fig:DLTS_P1_P3_P4_300C} shows DLTS spectra of samples irradiated to three different fluences, and subsequently annealed at \SI{300}{\degreeCelsius} (pre-diffusion) for measurement temperatures between about \SI{180}{\kelvin} and \SI{570}{\kelvin}, where six frequently reported trap levels are visible. The activation energies of these levels are summarized in Table~\ref{tab:defect_prop}, together with assignments to specific point defects made in the literature. The defect levels dominating the spectra, independent of irradiation fluence, are the so-called Z\lwr{1/2} level, which has been attributed to the ${(0/2-)}$ CTL of the carbon vacancy V\lwr{C} \cite{Alfieri.2005,Son.2012}, and the S center possessing the levels S\lwr{1} and S\lwr{2}, which have been identified as the ${(-/2-)}$ and (2$-$/3$-$) CTLs of the silicon vacancy V\lwr{Si} \cite{Bathen.2019}. We consequently interpret their concentration as equivalent to the content of V\lwr{C} and V\lwr{Si} \footnote{In the former case, since two electrons are emitted because of the negative-$U$ property of the Z\lwr{1/2} center, $[V\lwr{C}] = [Z\lwr{1/2}]/2$ is assumed.}. The feature labelled EH\lwr{6/7} at the high-temperature end of the spectra has been associated with the deeper lying (2+/+/0) CTLs of V\lwr{C} in different configurations \cite{Booker.2016}. Two more peaks are noteworthy: EH\lwr{4}, a rather broad and asymmetric peak spreading between about \SI{350}{\kelvin} and \SI{450}{\kelvin}, and EH\lwr{5}, which in the \SI{300}{\degreeCelsius} annealed samples appears only as a weak shoulder on the low-temperature flank of EH\lwr{6/7}. By using a high-resolution weighting function for the DLTS spectra, it becomes apparent that EH\lwr{4} possesses a substructure, and that three levels in total are needed to explain the considerable broadening in temperature this feature exhibits (inset of Fig.~\ref{fig:DLTS_P1_P3_P4_300C}). The chemical identity of the defects or defect complexes producing both EH\lwr{4} and EH\lwr{5} has not been clarified so far. 

\begin{table}
	\caption{Properties of defect levels documented in Fig.~\ref{fig:DLTS_P1_P3_P4_300C} (trap energy with respect to conduction band edge, as well as assignments to specific point defects from the literature).}
	\label{tab:defect_prop}
	\begin{tabular}{lcc}
	\toprule
	label & $E\mlwr{C}-E\mlwr{t}$ (\si{\electronvolt}) &  assignment, reference \\
	\midrule
	S\lwr{1} & \num{0.42} & V\lwr{Si} (2$-$/3$-$) \cite{Bathen.2019} \\
	Z\lwr{1/2} & \num{0.67} & V\lwr{C} (0/2$-$) \cite{Alfieri.2005,Son.2012} \\
	S\lwr{2} & \num{0.71} &  V\lwr{Si} (1$-$/2$-$) \cite{Bathen.2019} \\
	EH\lwr{4} & $\approx \num{1.0}$ &  \\
	EH\lwr{5} & $\approx \num{1.1}$ &  \\
	EH\lwr{6/7} & \num{1.5}-\num{1.6} &  V\lwr{C} (2+/+/0) \cite{Son.2012,Booker.2016} \\
	\bottomrule
	\end{tabular}
\end{table}

\begin{figure*}
	\centering
	\includegraphics[width=2\columnwidth]{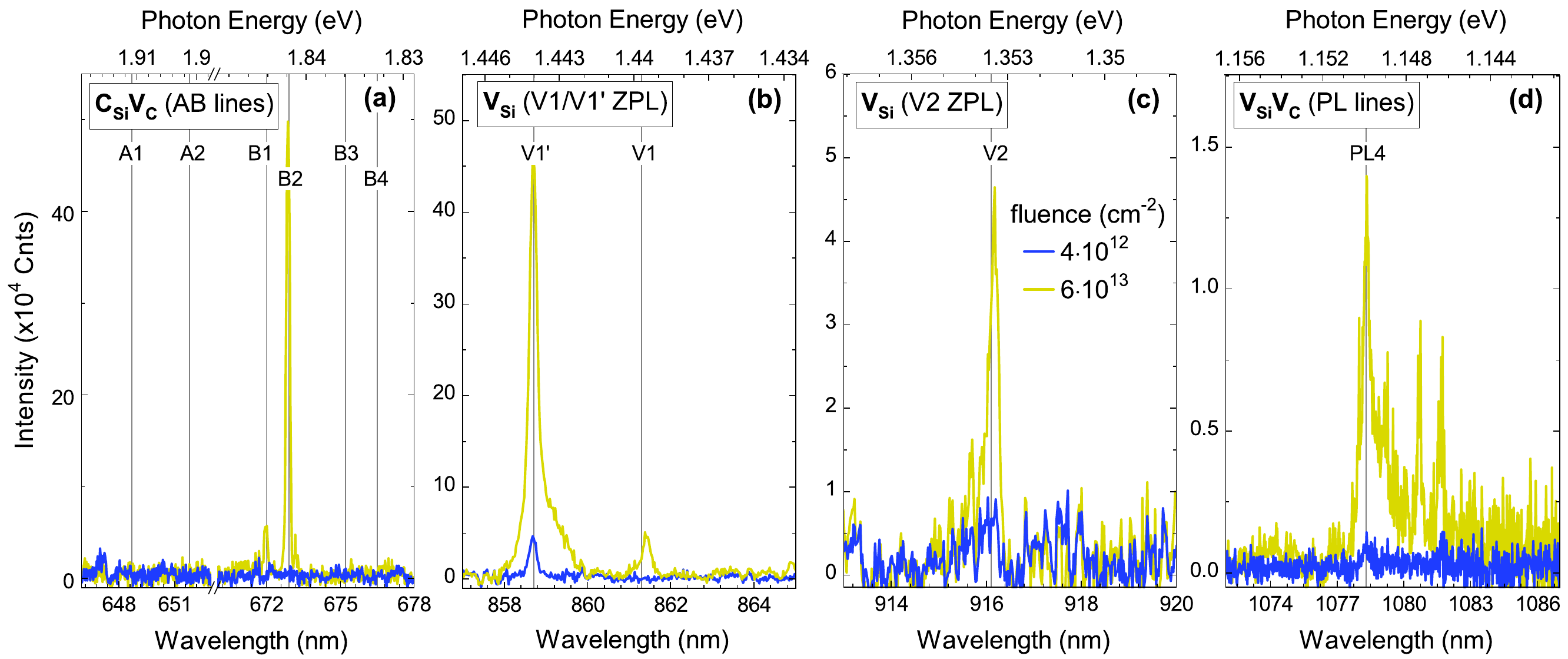}
	\caption{PL spectra of SiC epilayers irradiated to two different fluences and post-annealed at \SI{300}{\degreeCelsius}. Measurements were done at \SI{10}{\kelvin} in four different spectral windows containing the emission of (a) the CAV defect (AB lines), V\lwr{Si} with (b) its V1/V1' doublet and (c) V2 lines, and (d) the divacancy (PL4 line). Excitation at \SI{405}{\nano\meter}, \SI{75}{\milli\watt} (cw). Vertical grey lines are reference values for the lines.}
	\label{fig:PL_all_spectra_after300C}
\end{figure*}

While the V\lwr{C} acts mainly as a non-radiative recombination center and therefore does not emit light, V\lwr{Si}, CAV and VV have well-documented PL lines in certain charge states. Fig.~\ref{fig:PL_all_spectra_after300C} demonstrates the light emission from these three defects in two irradiated (to different fluences) samples at pre-diffusion conditions. Due to its energetically inequivalent configurations, the positively charged CAV pair (C\lwr{Si}V\lwr{C})\upr{+} exhibits a set of multiple emission lines in the range \SIrange[range-phrase=--]{648}{677}{\nano\meter} (labeled AB lines, Fig.~\ref{fig:PL_all_spectra_after300C}(a)) \cite{Steeds.2009}. As can be seen, AB emission becomes visible for increasing irradiation fluence, and is channeled mainly into the B1/B2 line pair associated with excited states of the $kk$ configuration of CAV \cite{Ivady.2014}. Importantly, it can be stated that the CAV pair is present in our samples to detectable amounts after proton irradiation and pre-diffusion.\\
Figs.~\ref{fig:PL_all_spectra_after300C}(b) and \ref{fig:PL_all_spectra_after300C}(c) display the emission windows from the silicon vacancy. The so-called V lines appear in two regions, one of which contains the double line V1/V1' at \num{858.7} and \SI{861.3}{\nano\meter}, and the other is the single V2 line at \SI{916.1}{\nano\meter}. These have been associated with emission from excited states of the single negative V\lwr{Si} defect on the $h$ and $k$ site for the V1/V1' and V2 lines, respectively \cite{Ivady.2017}. Under the experimental conditions used here, emission from V\uplwr{-}{Si} is mainly channeled into the V1' line.  \\
Emission from the divacancy V\lwr{Si}V\lwr{C}, in its neutral charge state, is shown in the fourth window in Fig.~\ref{fig:PL_all_spectra_after300C}(d). Analogously to the CAV pair, the VV is a two-component complex possessing four inequivalent configurations as well, each emitting at a specific IR wavelength in a wide range between \num{1077} and \SI{1132}{\nano\meter} \cite{Koehl.2011,Magnusson.2018}. The lines are labeled PL1, PL2, PL3 and PL4, of which we only show the latter due to detection limitations for longer wavelengths. PL4 has been shown to belong to the $hk$ configuration of the VV defect \cite{Magnusson.2018}. Similar to CAV, the divacancy is present in irradiated and pre-diffused SiC in our experiments.

\subsection{Identification of the C\lwr{Si}V\lwr{C} defect by DLTS}

\begin{figure}
	\centering
	\includegraphics[width=\columnwidth]{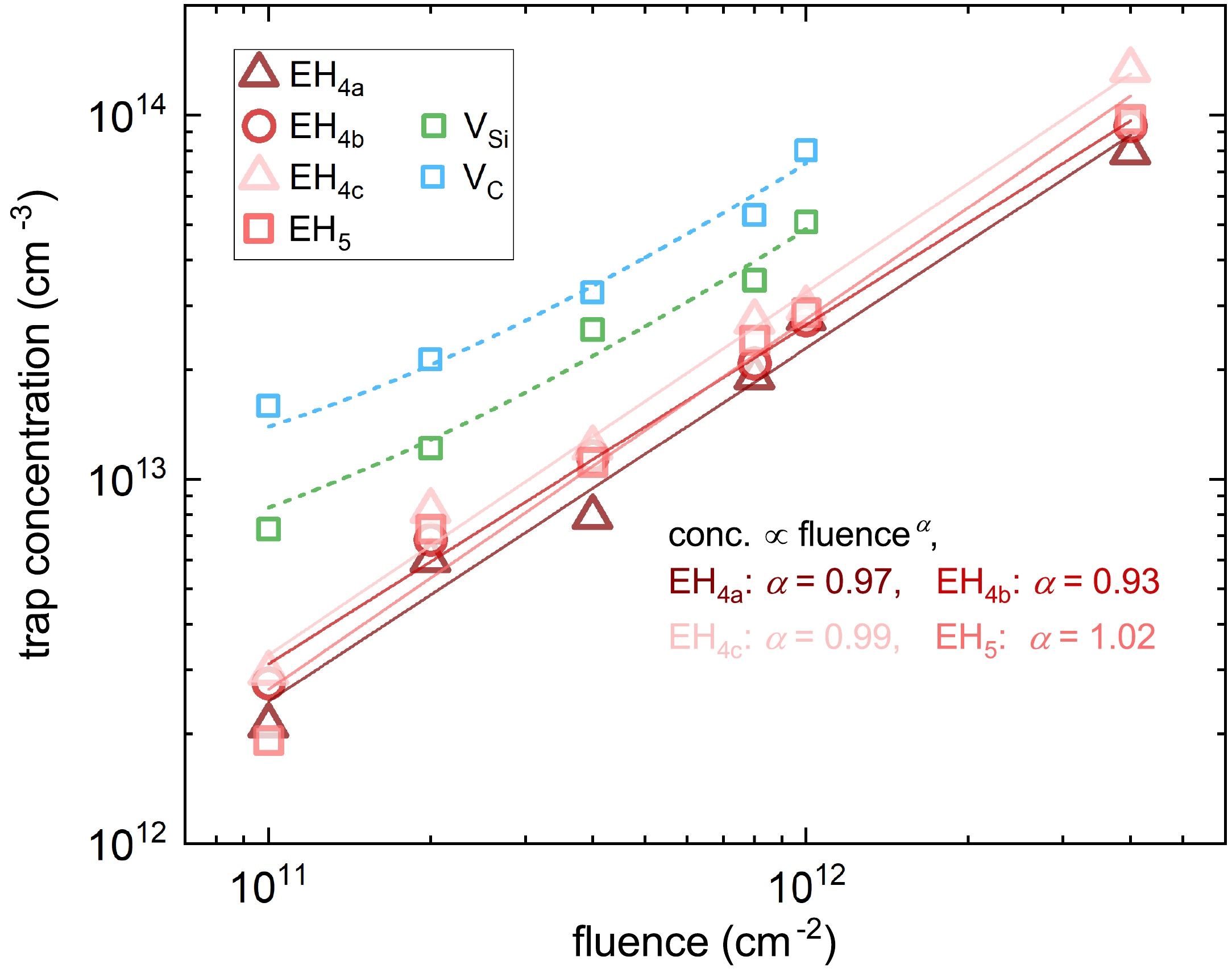}
	\caption{Fluence dependence of the concentration of EH4 and EH5 traps, as well as V\lwr{C} and V\lwr{Si}, after pre-diffusion. Compact lines: fits demonstrating the linear fluence dependence of EH\lwr{4,5}traps. Dashed lines: fits to V\lwr{C} and V\lwr{Si} concentrations according to a shifted-linear model.}
	\label{fig:EH45+VC+VSi_vs_fluence}
\end{figure}

\begin{figure}
	\centering
	\includegraphics[width=\columnwidth]{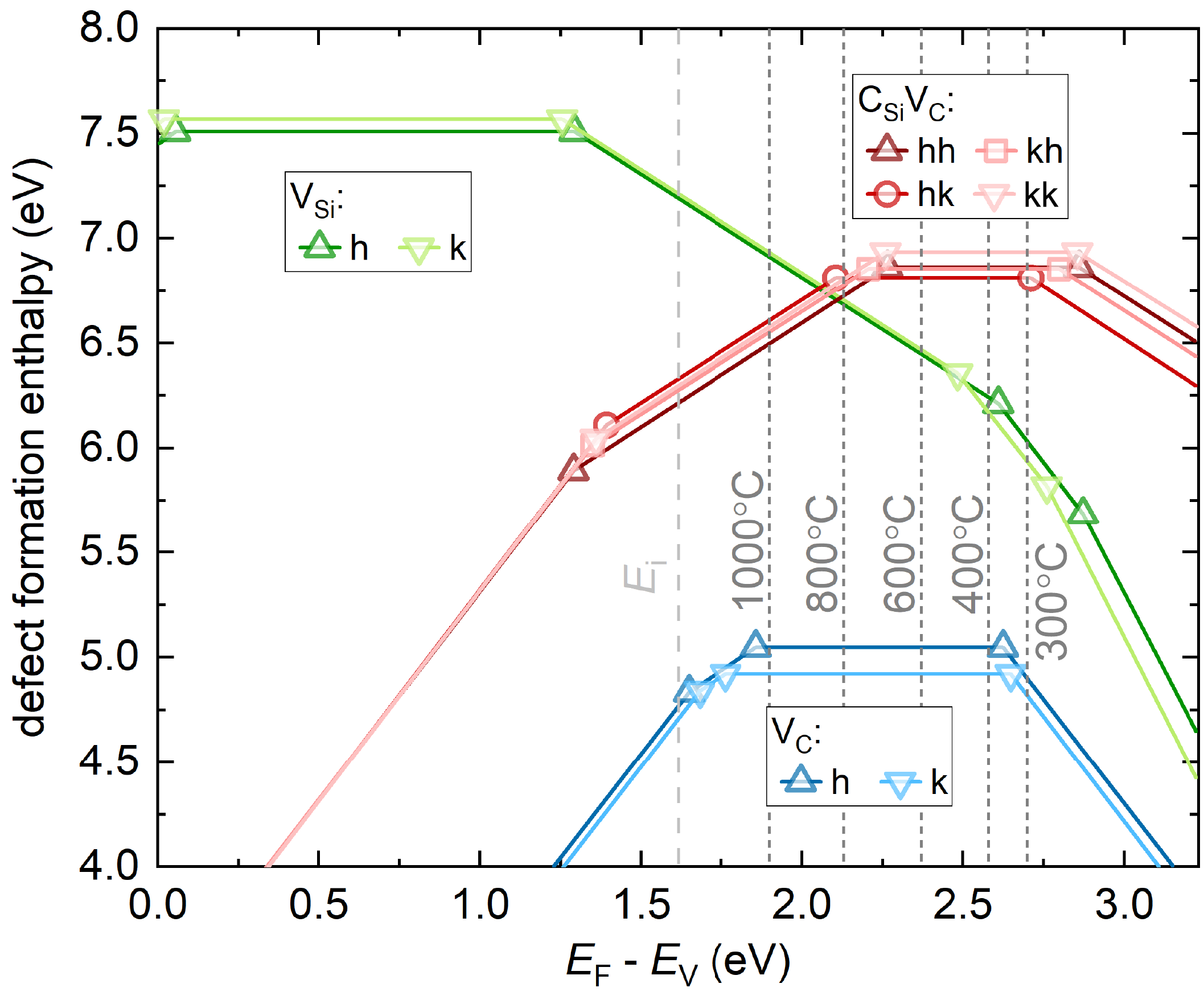}
	\caption{Defect formation enthalpies in the carbon-rich limit for the V\lwr{Si}, V\lwr{C} and C\lwr{Si}V\lwr{C} defects in different configurations, values adapted from Ref.~\cite{Szasz.2015}. Position of Fermi energy for the temperatures used in our annealing experiment, and intrinsic Fermi level present after high-fluence irradiation, indicated as dashed lines.}
	\label{fig:defect_formation_energies}
\end{figure}

\begin{figure}
	\centering
	\includegraphics[width=\columnwidth]{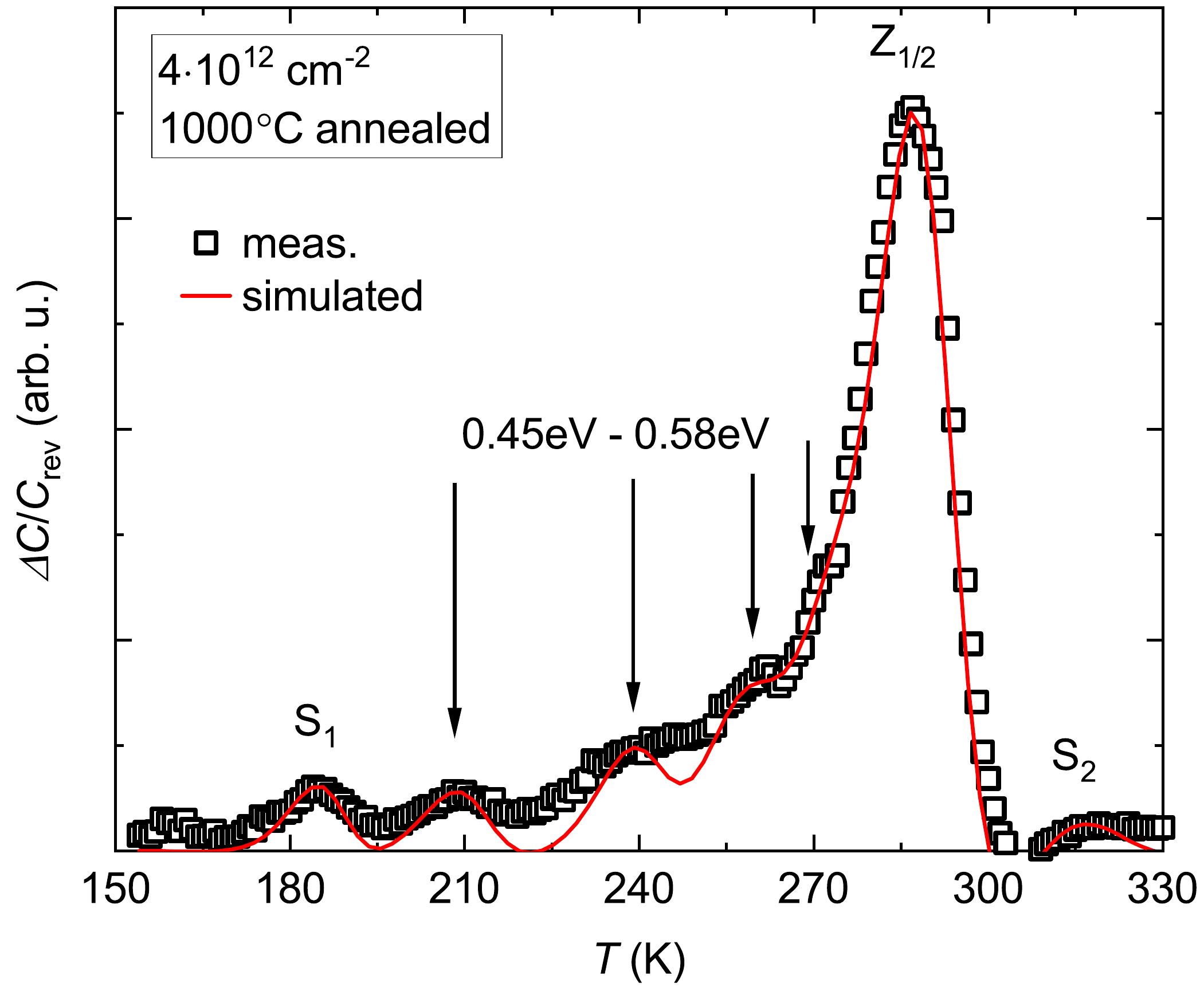}
	\caption{High-resolution DLTS spectrum of a high-fluence irradiated and \SI{1000}{\degreeCelsius} annealed sample, revealing a substructure of the low-temperature flank of Z\lwr{1/2} which is possibly related to the ${(0/-)}$ CTL of the CAV pair.}
	\label{fig:D7_after1000C_lowTemp+sim}
\end{figure}

Turning to the defect identity giving rise to the EH\lwr{4} and EH\lwr{5} traps in DLTS, we start with the observation that the occurence of these traps is correlated, i.e. either both or none of them are detectable in the same experiment. Moreover, their concentrations are always seen to be very similar, which is shown for the present study in Fig.~\ref{fig:EH45+VC+VSi_vs_fluence}. We take this as an indication for a common origin of the traps. Moreover, the EH\lwr{4,5} traps occur in concentrations similar to, but slightly lower than, those of the isolated vacancies V\lwr{C}, V\lwr{Si} (Fig.~\ref{fig:EH45+VC+VSi_vs_fluence}). For these defects, dynamic annealing and the subsequent \SI{300}{\degreeCelsius} anneal have been estimated to leave \SI{3}{\percent} of the initially created vacancies remaining \cite{Bathen.2019}. Based on their similar concentrations, it can be stated that the rate of introduction of the defect causing the EH\lwr{4,5} traps to appear is comparable, although slightly lower, than that of the monovacancies. This rules out larger defect complexes as the origin, a claim that is also supported by the fluence dependence of the EH\lwr{4,5} level concentration in Fig.~\ref{fig:EH45+VC+VSi_vs_fluence}. For a primary defect like V\lwr{Si} and V\lwr{C}, above a certain background level, a linear relationship of concentration with fluence is expected. This is exemplified by the linear fits\footnote{Here, the simple linear model $[\text{V}\lwr{C/Si}] = [\text{V}\lwr{C/Si}]^0 + i\mlwr{C/Si}\cdot d,$ was used where $[\text{V}\lwr{C/Si}]^0$ are the background concentrations, $i\mlwr{C/Si}$ are the defect introduction rates by irradiation and $d$ is the fluence.} to the monovacancy concentrations in Fig.~\ref{fig:EH45+VC+VSi_vs_fluence}. For higher-order defect complexes, the fluence dependence is expected to be superlinear, as multiple displacement events have to occur at adjacent lattice sites in order to create them. Fig.~\ref{fig:EH45+VC+VSi_vs_fluence} shows that for all four traps EH\lwr{4,5}, the fluence dependence is linear, which combined with the large introduction rate narrows down the list of candidates responsible for these levels to small complexes involving V\lwr{C}, V\lwr{Si}, Si\lwr{i} and C\lwr{i}, and possibly another impurity contained in the sample with significant concentrations. However, a participation of interstitials is unlikely due to their low thermal stability \cite{Bockstedte.2004} as compared to the EH\lwr{4,5} levels. This will be further supported by the thermal evolution of the defect population below. Moreover, electron irradiation experiments with energies below and above the Si displacement threshold \cite{Storasta.2004,Beyer.2012} have shown that EH\lwr{4,5} only appear when V\lwr{Si} is created. Therefore, the Si vacancy is likely to be involved in their formation.\\
Since the nitrogen content in the samples is approximately \SI{e15}{\per\cubic\centi\meter}, an additional involvement of N is conceivable, for example in the form of N\lwr{C}V\lwr{Si} (NV) centers. There are, however, arguments opposing this assignment. The formation of NV centers during room-temperature irradiation is possible when a V\lwr{Si} is formed adjacent to a N\lwr{C} donor. Because this process is much less likely than the formation of V\lwr{Si} at any lattice site, the initial density of NV centers is lower than [V\lwr{Si}] (brackets $[\;]$ denote concentration) by some orders of magnitude, so we do not expect NV to be detectable in DLTS in the pre-diffusion stage -- in particular not in concentrations comparable to [V\lwr{Si}] itself, as seen in Fig.~\ref{fig:EH45+VC+VSi_vs_fluence}. Further, the EH\lwr{4,5} levels have been observed in a variety of SiC material from different sources, including nominally undoped epilayers \cite{Hemmingsson.1997}. Consequently, the candidates responsible for the EH\lwr{4,5} traps can be further narrowed down to simple and intrinsic defect complexes involving V\lwr{Si}. This leaves mainly two possibilities: the CAV pair, and the divacancy.\\
The formation energy diagram for the V\lwr{C}, V\lwr{Si} and CAV defects as calculated in Ref.~\cite{Szasz.2015} is reproduced in Fig.~\ref{fig:defect_formation_energies}. CTLs related to the divacancy have not been calculated in the referenced work, but are generally expected to be in the range \SIrange[range-phrase=--]{0.7}{0.9}{\electronvolt} and \SIrange[range-phrase=--]{1.2}{1.4}{\electronvolt}, as discussed in the introduction. It must be stated that no prominent traps with activation energies in that range have been found by DLTS in the present work. Moreover, even though VV could be detected by PL for the highest-fluence samples in the pre-diffusion stage, [VV] is expected to be low in the lowest-fluence sample set when compared to  V\lwr{Si}, for example, while [S\lwr{2}] and [EH\lwr{4,5}] are of the same magnitude in these samples. We therefore argue for discarding the divacancy as the origin of the EH\lwr{4,5} traps. For the CAV pair, the formation energy diagram predicts the ${(+/0)}$ and the ${(0/-)}$ transitions to be at around \SI{1}{\electronvolt} and \SI{0.5}{\electronvolt} below $E\mlwr{C}$, respectively. While the latter is energetically similar to the ${(-/2-)}$ and (2$-$/3$-$) CTLs of the V\lwr{Si} (the S\lwr{1} and S\lwr{2} centers) and the ${(0/2-)}$ double acceptor transition of the V\lwr{C} (the Z\lwr{1/2} level), the deeper-lying ${(+/0)}$ CTL is energetically isolated and are therefore more suitable for the identification of the defect. An important observation of Fig.~\ref{fig:defect_formation_energies} is the splitting of the ${(+/0)}$ CTL belonging to the different CAV configurations into a set of three ($hk$, $hh$, $kk$) at around \SI{1.0}{\electronvolt} and a single one ($kh$) at \SI{1.13}{\electronvolt}. Referring to Table~\ref{tab:defect_prop}, it becomes evident that these agree well with the activation energies found for the EH\lwr{4} and EH\lwr{5} traps. As was discussed in the last section, the EH\lwr{4} peak consists of three sublevels (EH\lwr{4a,b,c}) with very similar activation energies of around \SI{1}{\electronvolt}, which is consistent with the hypothesis that EH\lwr{4} is produced by the $hk$, $hh$ and $kk$ configurations of the CAV pair, while EH\lwr{5} is connected to the slightly deeper $kh$. Indeed, a recent EPR study \cite{Son.2019} has found the ionization energies of the neutral CAV pair to be roughly \SI{1.1}{\electronvolt}, with detectable energetic splitting between the configurations, which further supports our assignment. \\
We next turn to the more elusive ${(0/-)}$ transition at lower activation energies. It is conspicuous that Z\lwr{1/2} possesses an extended low-temperature flank towards S\lwr{1}, rendering it asymmetric in shape. This flank, upon annealing at \SI{1000}{\degreeCelsius}, developed a clearly resolvable substructure containing four trap levels with activation energies in the range between \num{0.45} and \SI{0.58}{\electronvolt} (indicated by arrows in Fig.~\ref{fig:D7_after1000C_lowTemp+sim}), having signal intensities comparable to those of the EH\lwr{4/5} centers which were also present in this sample. The same substructure can be seen in DLTS spectra published by Alfieri \textit{et al.} (Fig.~5 in Ref.~\cite{Alfieri.2005}). These levels could therefore be the expected acceptor transitions of the CAV pair. 

\subsection{Evolution of defect concentrations with annealing}

\begin{figure}
	\centering
	\includegraphics[width=\columnwidth]{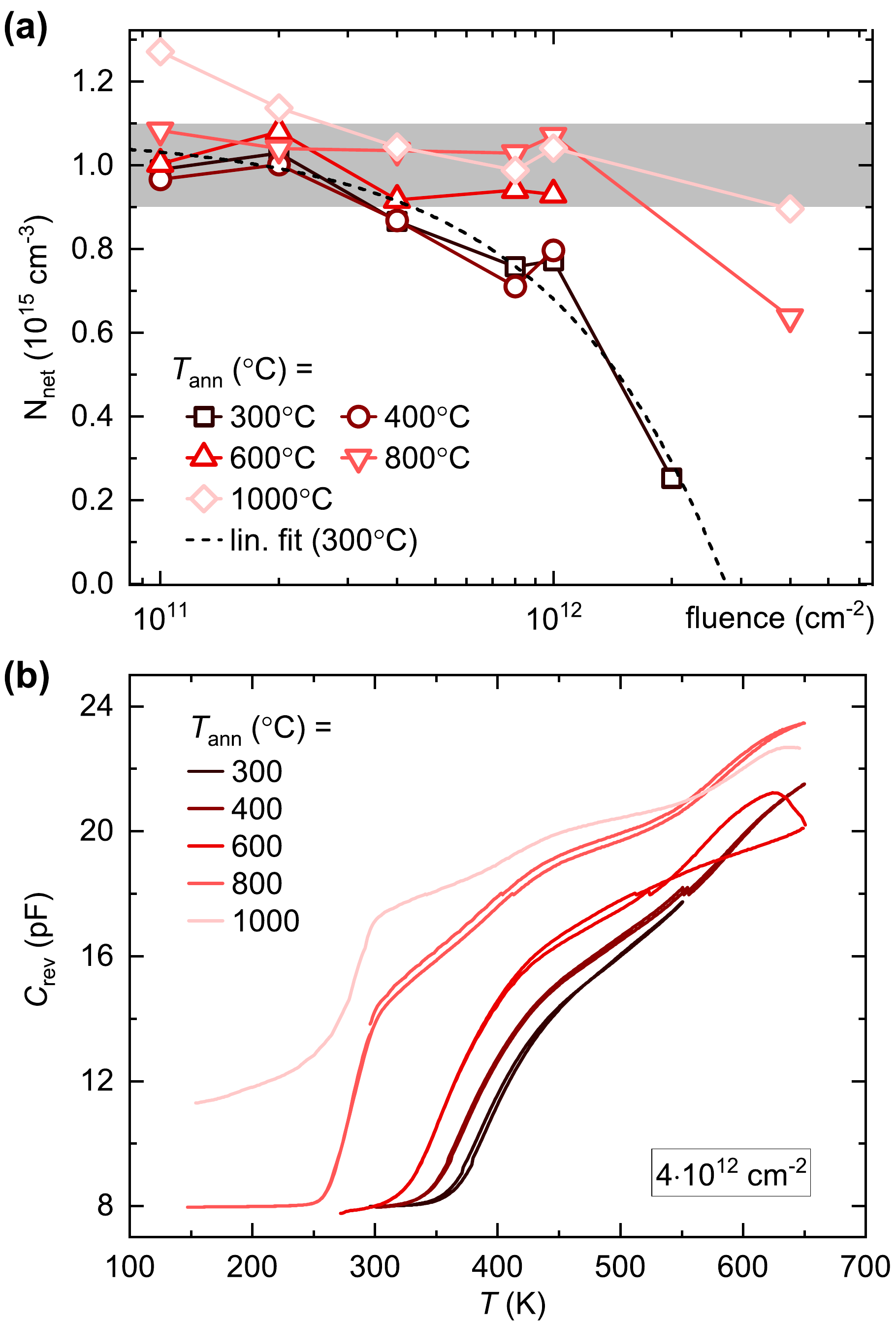}
	\caption{(a) Net doping concentrations of low-fluence irradiated samples after annealing at different temperatures; data collected by means of $C$-$V$ measurements at a reverse voltage of \SI{10}{\volt}. Dashed line: linear fit, dashed-dotted line: exponential fit to initial (\SI{300}{\degreeCelsius}) doping data. Nominal doping range highlighted in grey. (b) $C\mlwr{rev}$-$T$ curves of a sample exhibiting initial compensation and partial free carrier recovery upon heating (reversible; temperature sweeps in both directions are shown) and annealing (irreversible).}
	\label{fig:doping_conc_all_CT_D7}
\end{figure}

In this section, the different pathways through which intrinsic defects in $n$-type 4H-SiC can anneal out will be explored. The development of the defect population after each annealing step was monitored using DLTS for the low-fluence and PL for the high-fluence irradiated samples. Irradiation introduces a large variety of defects, and the dominant defects in 4H-SiC, like V\lwr{C}, V\lwr{Si} and the complexes VV and CAV have acceptor-like states in the upper part of the band gap. Therefore it is not surprising that such experiments usually lead to the removal of free carriers with increasing irradiation fluence \cite{Aberg.2001,Castaldini.2006,Kozlovskii.2008,Pastuovic.2017}. This was also observed in the present work. In Fig.~\ref{fig:doping_conc_all_CT_D7}(a) the net doping of the low-fluence samples, as determined by means of capacitance-voltage measurements, is shown after each annealing step. It can be seen that starting from a proton fluence of \SI{4e11}{\per\square\centi\meter} an initial reduction of net doping becomes noticeable. For the sample with the highest fluence in this sample set, room temperature {$C$-$V$} measurements were not possible in the initial state as the free carriers were entirely compensated. Interestingly, annealing can recover the free carriers almost entirely, with the temperature required to achieve recovery depending on the initial compensation. DLTS measurements on the highest-fluence sample across the full temperature range therefore became possible only after the \SI{800}{\degreeCelsius} anneal. However, even for the partially compensated state, gradual charge carrier recovery was observed for elevated temperatures, e.g. approximately \SI{400}{\degreeCelsius} for the initial (\SI{300}{\degreeCelsius}) stage, as is demonstrated in Fig.~\ref{fig:doping_conc_all_CT_D7}(b) where the temperature dependence of the reverse capacitance $C\mlwr{rev}$ is shown. The temperature at which this capacitance recovery occurs decreased after each annealing step. The recovery during measurements was shown to be reversible, in contrast to the effect induced by annealing.  This behavior is indicative of the compensation being due to low-stability acceptors that form independently of the doping concentration, and in contradiction to a previously suggested model according to which carrier removal is due to the irradiation-induced passivation of shallow N\lwr{C} donors \cite{Aberg.2001}. Introduction of compensating acceptors removes charge carriers in dependence on the Fermi level position which changes with temperature. Consequently, a partial recovery of free carriers at elevated temperatures is expected. The compensation hypothesis is also supported by the observation of the net doping density decreasing linearly with irradiation fluence, as is demonstrated by the fit (dashed line) in Fig.~\ref{fig:doping_conc_all_CT_D7}(a). Importantly, the total concentration of the detectable acceptor-like defects with known concentrations (V\lwr{C}, V\lwr{Si}, CAV pair) in our experiments is close to the concentration of the observed carrier removal. Hence the removal effect can to a large extent be explained by the aforementioned defects, indicating that there are no other prominent defects that have to be taken into account. \\
It is also noteworthy from Fig.~\ref{fig:doping_conc_all_CT_D7} that for the lowest irradiation fluences, annealing increases the net doping to values significantly above the nominal doping of \SI{e15}{\per\cubic\centi\meter}. We have currently no definite explanation for this behavior, but it is possible that the as-received SiC either already contained thermally unstable compensating acceptors that anneal out at around \SI{1000}{\degreeCelsius}, or a population of inactive donors that became activated after the anneal.

\subsubsection{Thermal evolution of the carbon vacancy}

\begin{figure}
	\centering
	\includegraphics[width=\columnwidth]{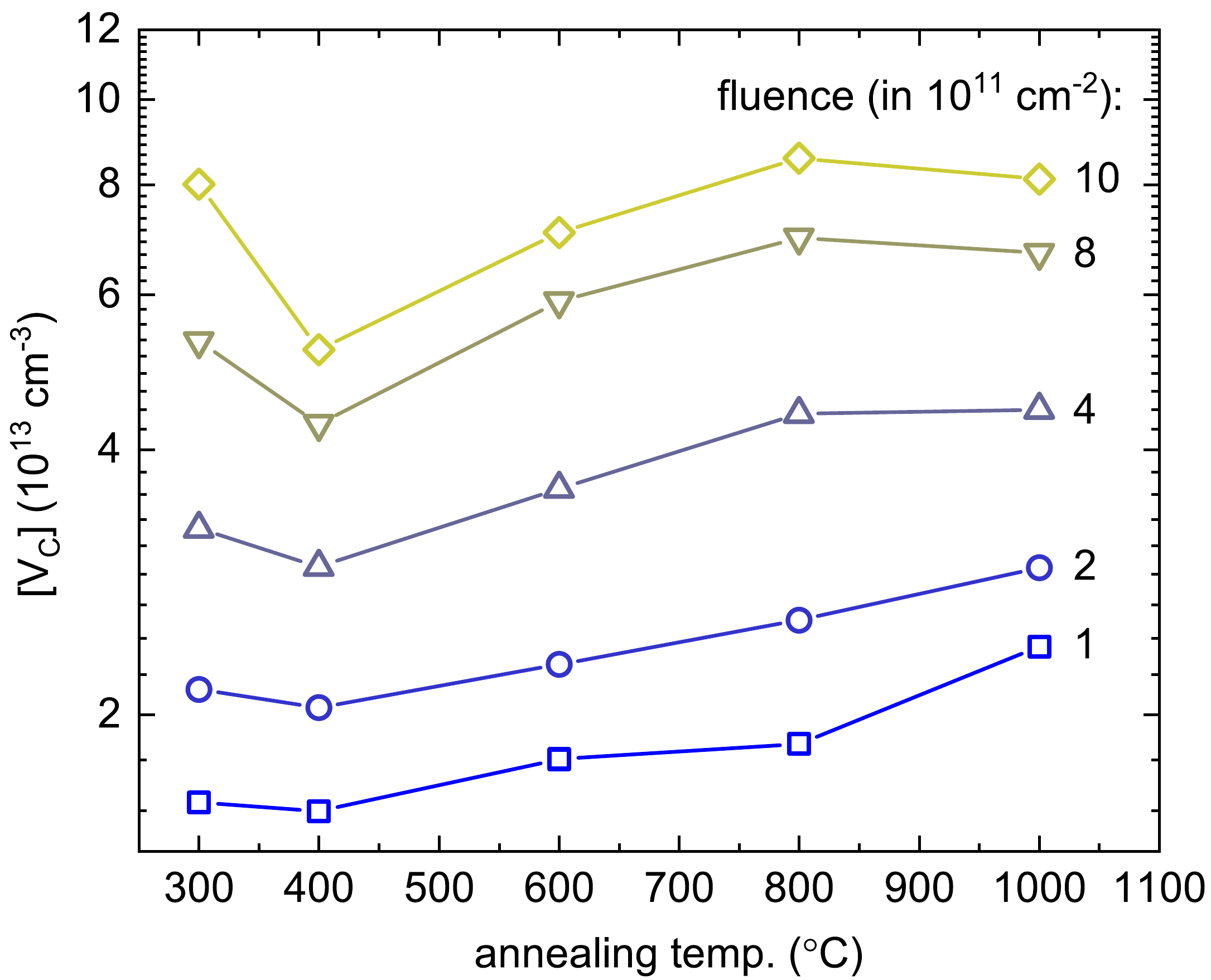}
	\caption{C vacancy concentrations, as determined from the Z\lwr{1/2} level intensity, of SiC irradiated to various proton fluences after isochronal (\SI{30}{\minute}) annealing steps to the given temperatures.}
	\label{fig:C_vac_conc_vs_annealing_temp}
\end{figure}

Fig.~\ref{fig:C_vac_conc_vs_annealing_temp} shows the development of the V\lwr{C} concentration, as determined from the DLTS signal of the Z\lwr{1/2} center, with annealing at increasingly higher temperatures for the different irradiation fluences. [V\lwr{C}] initially drops at \SI{400}{\degreeCelsius} and then increases with increasing annealing temperatures. The relative amount of the initial decrease is seen to be lowest for the lowest fluence, i.e. more V\lwr{C} is annealed out at low temperatures when the initial defect concentration is higher. For the highest annealing temperature of \SI{1000}{\degreeCelsius}, there is a renewed drop of [V\lwr{C}] for higher fluences. \\
A comparable observation was made by Alfieri \textit{et al.} \cite{Alfieri.2005} in a multi-stage annealing experiment on electron-irradiated 4H-SiC. Irradiation is known to produce self-interstitials and vacancies in comparable concentrations. Dynamic annealing during the irradiation and the subsequent \SI{300}{\degreeCelsius} pre-diffusion step are believed to lead to a significant decrease in interstitial concentration, but our results, combined with those by Alfieri \textit{et al.}, demonstrate that annealing at up to \SI{400}{\degreeCelsius} is necessary to produce a stable V\lwr{C} concentration by inducing recombination with residual interstitials. In fact, based on a model for vacancy-interstitial recombination developed by Bockstedte, Mattausch and Pankratov \cite{Bockstedte.2004}, the annihilation of these intrinsic defects is expected to occur in two stages in $n$-type SiC. Herein, the recombination of closely spaced vacancy-interstitial pairs happens at a comparatively low temperature of around \SI{200}{\degreeCelsius}, whereas pairs separated by more than the nearest-neighbor distance anneal out via an interstitial diffusion-limited process that requires slightly elevated temperatures of up to \SI{500}{\degreeCelsius}. Besides regular recombination which restore normally occupied lattice sites, formation of antisites through the same mechanism is also conceivable. However, Si\lwr{C} has been shown to only possess one CTL close to the valence band \cite{Kobayashi.2019} which is inaccessible in our experiments, such that it is not possible to distinguish antisite formation from C\lwr{i}-V\lwr{C} recombination. In any case, it can be stated that long-range interstitial diffusion is likely the reason for the continued annealing of V\lwr{C}. The process is enhanced under $n$-type conditions which produce defects in more negative charge states, lowering interstitial diffusion barriers \cite{Bockstedte.2003,Bockstedte.2004,Yan.2020}.\\
At the same time, considering the similarity of our experiments with those by Alfieri \textit{et al.}, it can be stated that the renewed increase of V\lwr{C} concentration above \SI{400}{\degreeCelsius} is a process that is induced under different experimental conditions (electron vs. proton irradiation). A conceivable explanation is thermal release of V\lwr{C} from defect complexes. It was previously shown in Fig.~\ref{fig:PL_all_spectra_after300C}(d) that the VV defect is actually present in the samples after the pre-diffusion step. The thermal dissociation of VV into the isolated vacancies would explain an increase of [V\lwr{C}]. However, the VV center has been shown to be thermally stable and have a high binding energy \cite{Torpo.2002,Son.2006,Xu.2009}, such that thermal dissociation below \SI{1100}{\degreeCelsius} is unlikely. Another possibility of V\lwr{C} release is from the CAV pair via its dissociation, which is discussed in Subsection 3. Analogously to VV, CAV has formed in detectable amounts over the course of dynamic annealing during proton irradiation, as has already been shown in Fig.~\ref{fig:PL_all_spectra_after300C}(a). 

\subsubsection{The silicon vacancy and the CAV pair}

\begin{figure}
	\centering
	\includegraphics[width=\columnwidth]{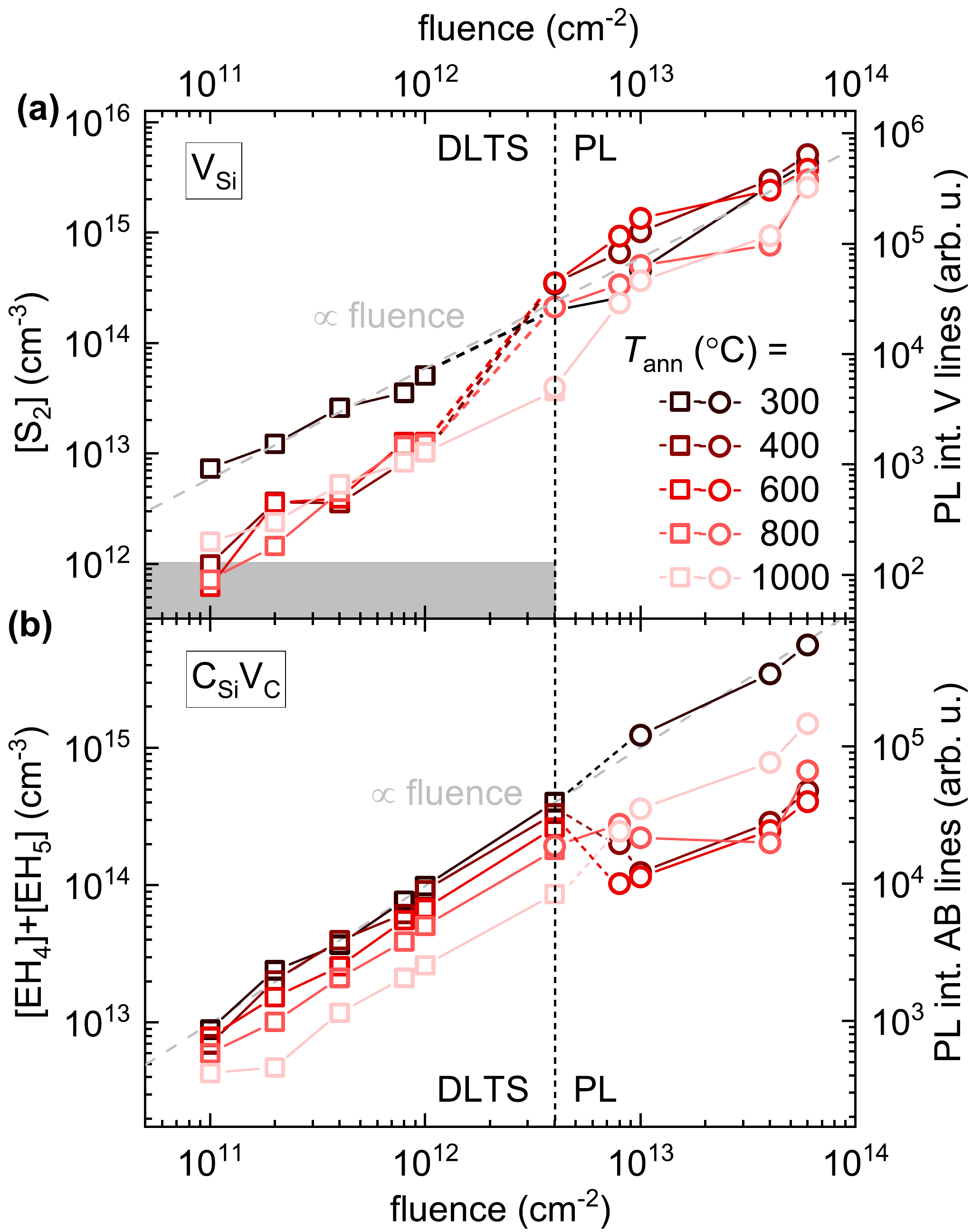}
	\caption{Fluence dependence of the (a) V\lwr{Si} and (b) CAV pair concentration, obtained from combined DLTS (S\lwr{2} and EH\lwr{4,5} traps) and PL results (V and AB lines) after annealing at increasingly high temperatures. The sample irradiated to a fluence of \SI{4e12}{\per\square\centi\meter} (dashed black line) could be characterized by both methods in several annealing stages and was therefore used for calibration. Grey area marks onset of increased inaccuracy in DLTS measurement.}
	\label{fig:VSi_CAV_conc_vs_fluence_DLTS_PL}
\end{figure}

PL spectroscopy is a technique used for the identification and relative quantification of defect concentrations, but not their absolute values. However, because the signatures of V\lwr{Si} and CAV in both the DLTS measurements (S center, EH\lwr{4,5 } traps) and the PL spectra (V lines, AB lines) have been identified, and the DLTS and PL results overlap for one fluence (\SI{4e12}{\per\square\centi\meter}), we normalized the PL data with regard to the DLTS concentrations in the overlapping region, and then used the established fixed multiplier to calibrate the rest of the PL data to obtain estimates for the absolute defect concentrations for both defects across the whole fluence range.\\
Fig.~\ref{fig:VSi_CAV_conc_vs_fluence_DLTS_PL}(a) shows that there is a distinct difference between the thermal evolution of the Si vacancy at low and at high irradiation fluences. For samples with fluences at and below \SI{1e12}{\per\square\centi\meter}, i.e. the dilute concentration regime, the evolution is dominated by an abrupt drop of [V\lwr{Si}] for anneals at \SI{400}{\degreeCelsius}. While annealing to increasingly higher temperatures, the V\lwr{Si} concentration remains almost constant. Analogously to the V\lwr{C} defect, it can be stated that the stark drop is probably due to recombination of residual Si and C interstitials that initially lack an immediate V\lwr{Si} neighbor (Si\lwr{i}-V\lwr{Si} recombination cannot be distinguished from C\lwr{Si} antisite formation by DLTS, as this defect does not have any CTLs in the band gap \cite{Kobayashi.2019}). This requires somewhat elevated temperatures due to longer diffusion ranges. The migration barrier for C\lwr{i} and Si\lwr{i} are below \SI{1}{\electronvolt} for the electrically neutral charge state and increase to around \SI{2}{\electronvolt} as the Fermi level moves deeper into the band gap\cite{Yan.2020}. This means that interstitial diffusion is facilitated in $n$-type and partially blocked for compensated material, which is a likely explanation for the absence of a decrease in concentration in the \SI{400}{\degreeCelsius} samples for fluences larger than \SI{e12}{\per\square\centi\meter}. For these compensated samples, the V\lwr{Si} concentration evolution is instead mainly characerized by a slow decrease above \SI{400}{\degreeCelsius}.  \\
To understand the evolution of V\lwr{Si}, it is insightful to compare it with the simultaneous development of the CAV pair concentration (Fig.~\ref{fig:VSi_CAV_conc_vs_fluence_DLTS_PL}(b)). Here, a similar dichotomy of the annealing behavior can be stated. At fluences of up to \SI{4e12}{\per\square\centi\meter}, [CAV] decreases slowly with annealing temperature, by a factor of \num{3.75} between \SI{300}{\degreeCelsius} and \SI{1000}{\degreeCelsius} independent of fluence; the decrease appears to be somewhat accelerated at the highest temperatures. For the higher fluences, within the PL sample set, the \SI{400}{\degreeCelsius} anneal induces a pronounced drop of [CAV] starting from a fluence of \SI{8e12}{\per\square\centi\meter} which is completely alleviated in the \SI{1000}{\degreeCelsius} step. \\
We conjecture that the evolution of both defects at fluences higher than \SI{4e12}{\per\cubic\centi\meter} are partially related to the interconversion of V\lwr{Si} and the CAV pair. The preferential direction of this process depends on the position of the Fermi level (see formation energy diagram, Fig.~\ref{fig:defect_formation_energies}): in $p$-type and intrinsic SiC, the CAV pair possesses lower formation energies than V\lwr{Si} and will therefore be the more favorable defect species, while the situation is reversed for $n$-type conditions when the Fermi energy is closer to the conduction band than approximately \SI{1}{\electronvolt}. Additionally, the transformation between those defects is connected to a migration barrier $E\mlwr{m}$ for the C atom which also depends on the position of the Fermi level. Further, since for a fixed $E\mlwr{F}$, both V\lwr{Si} and CAV are never in the same charge state (see Fig.~\ref{fig:defect_formation_energies}), there are electron capture or emission processes associated with the conversion. For the charge states most likely occupied in our experiments, and considering only those reactions that yield an overall energy gain upon transformation, those processes read

	\begin{align}
		\text{V}\lwr{Si}\mupr{1-} + \text{C}\lwr{C} &\longrightarrow  \text{C}\lwr{Si}\text{V}\lwr{C}\mupr{1+} + 2e\mupr{-}, \; \label{eq:VSi_to_CAV} \\
		&(E\mlwr{m} = \SI{2.5}{\electronvolt}),  \nonumber 
	\end{align}

\begin{subequations}
	\begin{align}
		\text{C}\lwr{Si}\text{V}\lwr{C}\mupr{0} + n e\mupr{-} &\longrightarrow V\lwr{Si}\mupr{\textit{n}-} + \text{C}\lwr{C} \;  \label{subeq:CAV_to_VSi_2} \\
		 &(E\mlwr{m} = \SI{3.5}{\electronvolt}, n = 1,2)   \nonumber \\
		\text{C}\lwr{Si}\text{V}\lwr{C}\mupr{1+} + 2e\mupr{-} &\longrightarrow V\lwr{Si}\mupr{-} + \text{C}\lwr{C} \; \label{subeq:CAV_to_VSi_3}  \\
		&(E\mlwr{m} = \SI{4.2}{\electronvolt} \; \text{to} \; \SI{4.7}{\electronvolt})  \nonumber 
	\end{align}
	\label{eq:CAV_to_VSi}
\end{subequations}

with $n$ being the number of captured electrons, and the migration barriers $E\mlwr{m}$ from Ref.~\cite{Defo.2018} given in parentheses. The conversion of V\lwr{Si} to CAV (Eq.~(\ref{eq:VSi_to_CAV})) is hindered by a migration barrier of \SI{2.5}{\electronvolt}, but also includes double electron emission which requires an additional \SI{1.6}{\electronvolt} (\SI{0.5}{\electronvolt} for ionization of the first and \SI{1.1}{\electronvolt} for the second electron, based on the CAV pair CTLs), consequently increasing the transformation barrier to approximately \SI{4.1}{\electronvolt}. The reverse process, CAV to V\lwr{Si}, only requires free electrons available for capture, but the migration barriers are \SI{3.5}{\electronvolt} and higher. Since both reactions described by the equations (\ref{eq:VSi_to_CAV}) and (\ref{eq:CAV_to_VSi}) contain a single atomic hop, the corresponding hopping rate for a specific C atom at any given temperature is given by $f\mlwr{hop} = \nu\mlwr{0}\exp\left(\nicefrac{-E\mlwr{m}}{k\mlwr{B}T}\right)$, with $\nu\mlwr{0}$ the attempt frequency typically in the order of a typical phonon frequency, i.e. $\nu\mlwr{ph} \approx \SI{e13}{\per\second}$. Both V\lwr{Si} and V\lwr{C} are tetrahedrally coordinated, therefore the total fraction of V\lwr{Si} converted to CAV during an annealing experiment of duration $t$ can be approximated by

\begin{equation}
	\frac{[\text{C\lwr{Si}V\lwr{C}}](t)}{[\text{V\lwr{Si}}](t)} = 4t\cdot \nu\mlwr{ph} \exp \left(-\frac{E\mlwr{m}}{k\mlwr{B}T}\right),
	\label{eq:VSi_to_CAV_conversion_frac}
\end{equation}

and with inverted roles of V\lwr{Si} and CAV for the reverse process. Based on the above migration barrier of at least \SI{3.5}{\electronvolt} for the transformation of CAV to V\lwr{Si}, a considerable fraction of converted defects, say $\nicefrac{[\text{CAV}]}{\text{[V\lwr{Si}]}} \approx 0.5$, can be achieved above \SI{750}{\degreeCelsius}. However, the charge state of the CAV pair changes from neutral to 1+ just below \SI{800}{\degreeCelsius} (compare Fig.~\ref{fig:defect_formation_energies}). The associated CAV to V\lwr{Si} transformation barrier then increases to at least \SI{4.2}{\electronvolt} which shifts the required temperatures for conversion up even further. Given the fact that from \SI{800}{\degreeCelsius} on, the CAV pair also becomes energetically more favorable than V\lwr{Si}, we believe that the transformation CAV to V\lwr{Si} plays an insignificant role in $n$-type 4H-SiC. Conversely, using a combined migration and emission barrier of \SI{4.1}{\electronvolt} for the V\lwr{Si}-to-CAV process, this should be observable for the highest temperatures used in this experiment. There is no corresponding trend visible in the low-fluence samples, likely because a large fraction of V\lwr{Si} have recombined with interstitials until \SI{400}{\degreeCelsius}, and the remaining V\lwr{Si} concentration is an order of magnitude lower than that of the CAV pair. If the reaction described by Eq.~(\ref{eq:VSi_to_CAV}) occurs, its effect can be considered negligible compared to the already existing CAV defect population. In the high-fluence (compensated) samples however, where interstitial diffusion is suppressed, it should be observable. Indeed, the partial recovery of [CAV] for the highest temperatures may be due to this process. It is also accompanied by a decrease of [V\lwr{Si}] of comparable magnitude, as expected. 

\subsubsection{Correlation between CAV and V\lwr{C} concentrations}
\label{subsubsec:CAV+VC-correl}

\begin{figure}
	\centering
	\includegraphics[width=\columnwidth]{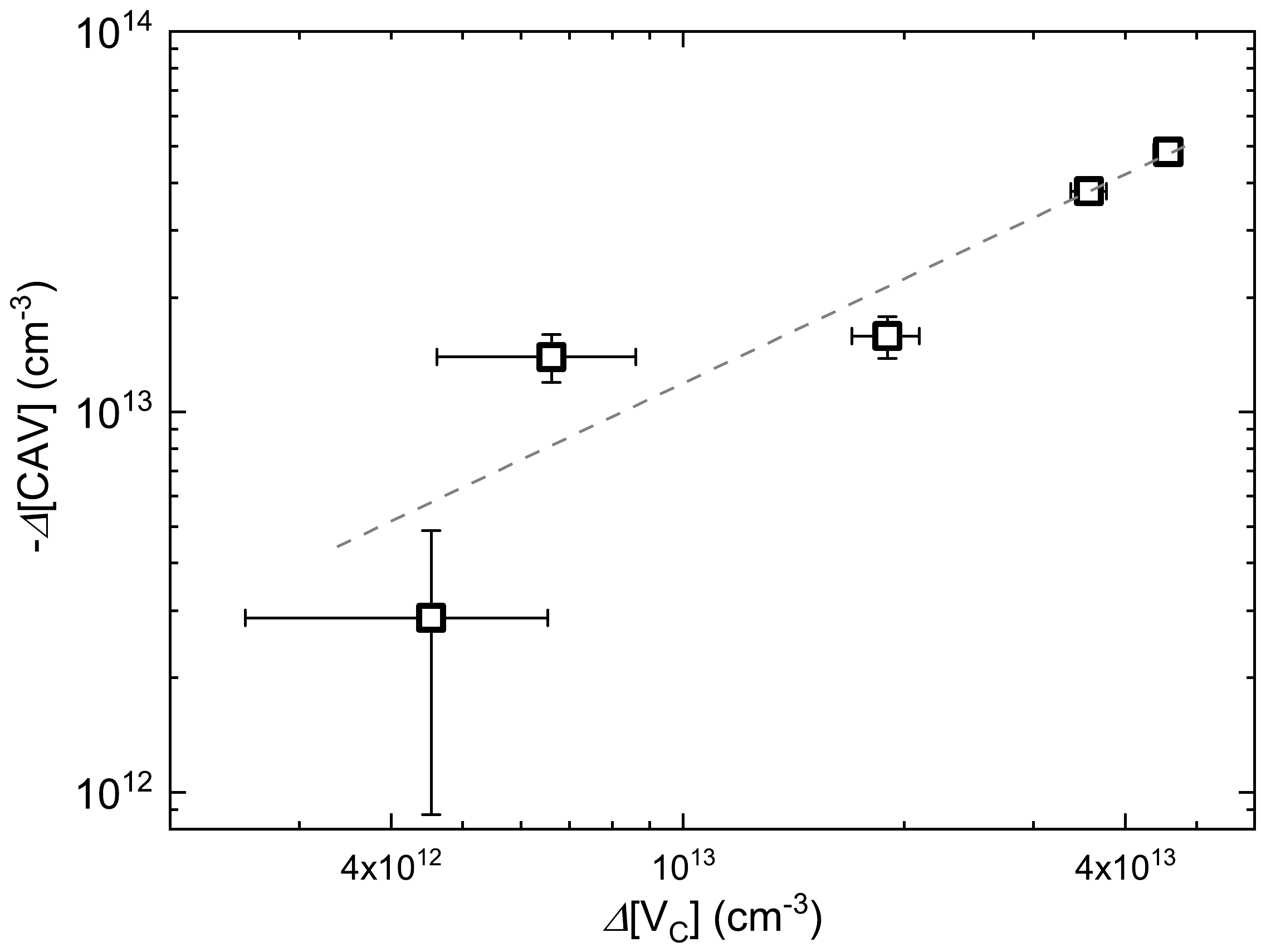}
	\caption{Correlation of concentration differences between V\lwr{C} and CAV defects. Grey dashed line indicates a linear 1:1 relation.}
	\label{fig:DeltaC_VC_CAV_vs_fluence}
\end{figure}

What remains to be explained is the decrease of [CAV] upon annealing, in particular the remarkable drop already at quite low temperatures for fluences of \SI{4e12}{\per\square\centi\meter} and higher. In the low-fluence range, [CAV] decreases by a factor of \num{3.75} in the course of the entire annealing experiment, independent of irradiation fluence. This indicates that the mechanism behind the decrease does not include other defect species. A mechanism which would conform to this restriction is the dissociation of CAV, as was already suggested earlier when the enhanced concentration of V\lwr{C} defects was discussed. The reaction $\text{C}\lwr{Si}\text{V}\lwr{C} \longrightarrow \text{C}\lwr{Si}  + \text{V}\lwr{C}$ produces an isolated carbon vacancy and a carbon on silicon-antisite C\lwr{Si}. It requires at least a single hop of a nearest-neighbor C atom into the vacancy in order to separate the two components. C\lwr{Si} is a defect without CTLs inside the band gap of 4H-SiC \cite{Torpo.2001,Kobayashi.2019}, and is therefore electrically inactive. For that reason, the dissociation of CAV pairs exclusively leads to the increase of V\lwr{C} concentration, as it was observed in our experiments (Fig.~\ref{fig:C_vac_conc_vs_annealing_temp}).  
Plotting the measured decrease of CAV pair concentration in the low-fluence samples with annealing temperature above \SI{400}{\degreeCelsius} versus the simultaneous increase of V\lwr{C} content from Fig.~\ref{fig:C_vac_conc_vs_annealing_temp}, it becomes apparent that these quantities agree well which each other, and share a linear 1:1 relation within experimental error (see Fig.~\ref{fig:DeltaC_VC_CAV_vs_fluence}). We interpret this as an indication that the decay of the concentration of CAV is closely related to the gain in V\lwr{C} density, or even the dominant contributor to this process. Using an expression similar to Eq.~(\ref{eq:VSi_to_CAV_conversion_frac}) for the estimation of the dissociation fraction, and assuming that the temperature of \SI{400}{\degreeCelsius} corresponding to the renewed increase of [V\lwr{C}] also marks the onset of CAV dissociation, the barrier associated with that process can be expected to be in the range of \SI{2.4}{\electronvolt}. However, there are not many theoretical considerations of the energetics of this process for comparison; one of the available works has been published by Wang \textit{et al.} \cite{Wang.2013} who consider the dissociation as one possible conversion channel of the CAV pair. Based on a comparatively high dissociation barrier, which they estimate to be \SI{5.9}{\electronvolt} and moreover identical to the migration barrier of V\lwr{C}, their assessment is that this process is not of significance compared to the inter-conversion between CAV and V\lwr{Si}; however, this assessment was made assuming $p$-type conditions. As was already stated, shifting the Fermi level towards $n$-type conditions can drastically change migration barriers for defects. In fact, for the neutral C vacancy, a migration barrier between \SI{3.7}{\electronvolt} and \SI{4.2}{\electronvolt} \cite{Bathen.2019b} has been calculated, which is considerably lower than the value given by Wang \textit{et al.}. However, this value would still be too large for the V\lwr{C} migration to play a significant role below \SI{1000}{\celsius}. \\
It can be debated whether the migration barrier for the single vacancy is actually a good approximation for the dissociation energy of the CAV in general. Although the process resembles V\lwr{C} migration, the energetics are different: both initial and final defect configuration of the dissociation (C\lwr{Si}V\lwr{C} vs. C\lwr{Si} $+$ V\lwr{C}) have higher formation energies compared to pure V\lwr{C} diffusion. For most annealing temperatures in our experiments, the CAV pair is in the neutral state, which according to Fig.~\ref{fig:defect_formation_energies} has a formation energy of roughly \SI{7}{\electronvolt}, about \SI{2}{\electronvolt} higher than the neutral state of V\lwr{C}. The final state consists of the separate V\uplwr{0}{C} and C\uplwr{0}{Si} defects, having formation energies of \SI{5}{\electronvolt} and roughly \SI{2}{\electronvolt} \cite{Kobayashi.2019}. Therefore, both initial and final state are about \SI{2}{\electronvolt} higher in energy than V\lwr{C}. If, apart from these constant energy shifts, the hopping carbon atom experiences the same electrostatic forces during its migration, it is conceivable that the dissociation barrier for the CAV pair is considerably lower (by an estimated 1 to \SI{2}{\electronvolt}) than the pure V\lwr{C} migration barrier. Taking into account the result of Wang \textit{et al.} \cite{Wang.2013} claiming that the dissociation energy of the CAV pair is identical for the case in which both components are second-nearest neighbors as well as the case where they are located inside different unit cells, it can also be stated that a single carbon hop is probably enough to restore the electronic properties of the isolated V\lwr{C}.\\
As for the compensated samples, it is presently unclear if the pronounced drop of [CAV] can be similarly attributed to the dissociation process. If it was, there would be a fundamental difference between the compensated and the $n$-type samples with regard to the annealing temperature that is necessary to induce the dissociation, with a considerably lower associated barrier for lower-lying $E\mlwr{F}$.

\subsubsection{The divacancy}

\begin{figure}
	\centering
	\includegraphics[width=\columnwidth]{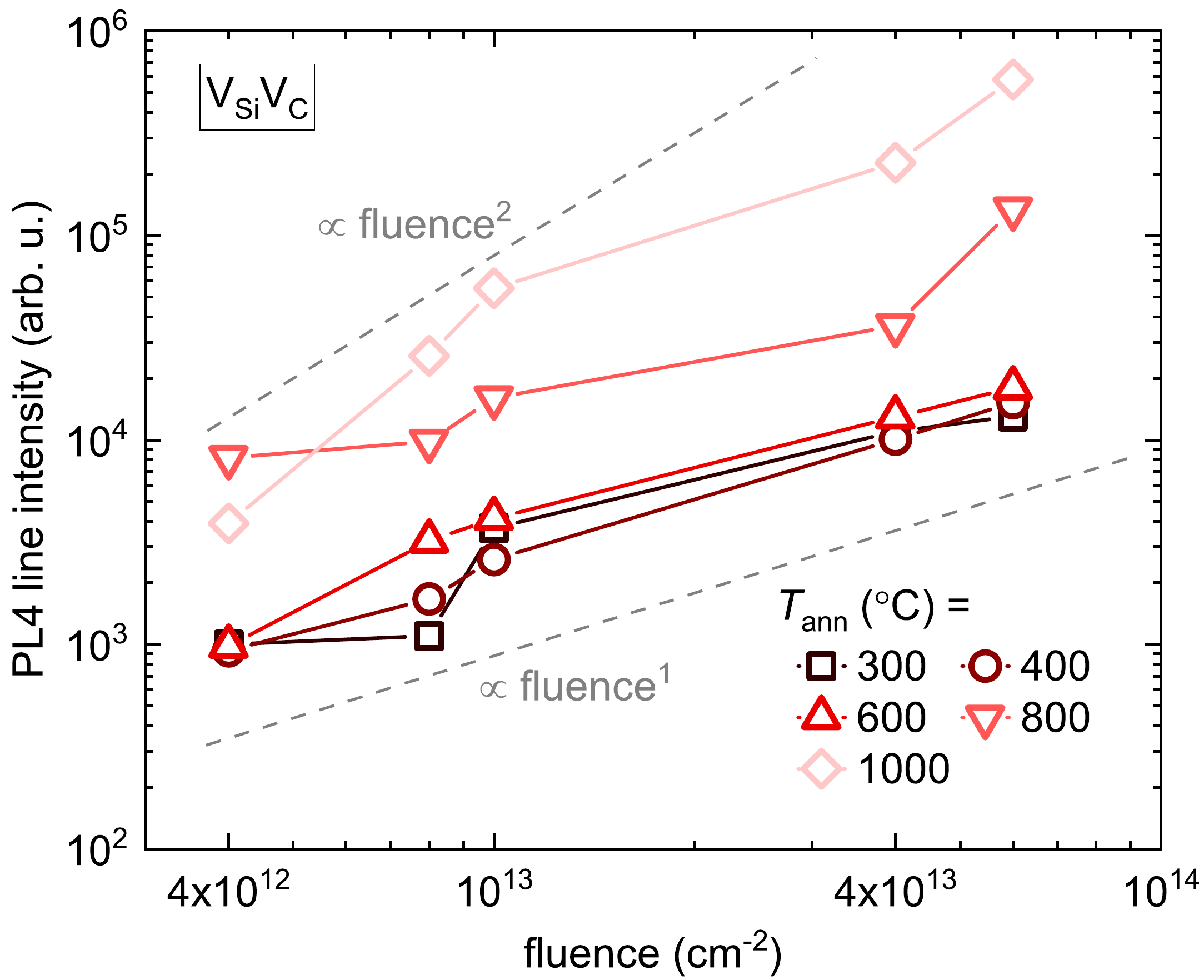}
	\caption{Evolution of the intensity of the PL4 line (divancancy V\lwr{Si}V\lwr{C}) with annealing temperature.}
	\label{fig:VV_vs_fluence_PL}
\end{figure}

The divacancy can form during irradiation, or at elevated temperatures during post-irradiation annealing when the isolated vacancies become mobile -- mostly due to V\lwr{Si} diffusion, because the carbon vacancy has been shown to be immobile below \SI{1200}{\degreeCelsius} \cite{Bathen.2019b}. Fig.~\ref{fig:VV_vs_fluence_PL} displays the development of the PL4 line intensity with fluence and annealing temperature, which is taken as a qualitative measure for [VV] here. After irradiation and the \SI{300}{\degreeCelsius} step, the divacancy is detectable in all high-fluence samples, following a linear relationship with fluence, as expected. Heating the samples to up to \SI{600}{\degreeCelsius} only leads to a minor change of [VV]; above that temperature, the divacancy concentration exhibits a steep increase, with the highest concentrations detected after \SI{1000}{\degreeCelsius}. [VV] now also follows an approximately quadradic relation with fluence, which can be attributed to the fact that a large portion of the defects have been formed by diffusion of V\lwr{Si} and their pairing with V\lwr{C}, the concentration of both of which is roughly proportional to the fluence. Son \textit{et al.} \cite{Son.2006} have reported on the formation of the EPR P6/7 center, which has likewise been attributed to the divacancy, in considerable amounts by annealing above \SI{750}{\degreeCelsius}, which matches the critical VV formation temperature in this study. \\
The formation of VV reduces both the V\lwr{Si} and V\lwr{C} density to a certain extent. The resulting [VV] cannot be directly inferred from the PL measurements. However, it can be taken from Figs.~\ref{fig:C_vac_conc_vs_annealing_temp} that the trend of a slow increase of V\lwr{C} density with temperature reverses at the highest annealing temperatures for irradiation fluences above \SI{2e11}{\per\square\centi\meter}. In contrast to the lower-fluence samples, the V\lwr{C}-V\lwr{C} inter-defect distance for the higher fluences becomes low enough for the V\lwr{Si} to migrate during annealing such that VV can form. A similar trend can be found for [V\lwr{Si}] (Fig.~\ref{fig:VSi_CAV_conc_vs_fluence_DLTS_PL}(a)): the \SI{8e11}{\per\square\centi\meter} and \SI{1e13}{\per\square\centi\meter} fluences exhibit a slight decrease of [V\lwr{Si}] at the highest annealing temperatures. For both defects, the change of concentrations when going from \SI{800}{\degreeCelsius} to \SI{1000}{\degreeCelsius} match well, and amount to approximately \SI{2e12}{\per\cubic\centi\meter} which would then correspond to the final divacancy concentration in these samples after the \SI{1000}{\degreeCelsius} anneal. This is an order of magnitude lower than the concentration of the CAV pair, for instance, which has CTLs in the same range as expected for the VV. It may therefore be that DLTS peaks corresponding to the VV defect are completely superimposed by the EH\lwr{4,5} traps, and hence are not resolvable in the spectra.

\subsection{Summary of defect evolution model}

\begin{figure}
	\centering
	\includegraphics[width=\columnwidth]{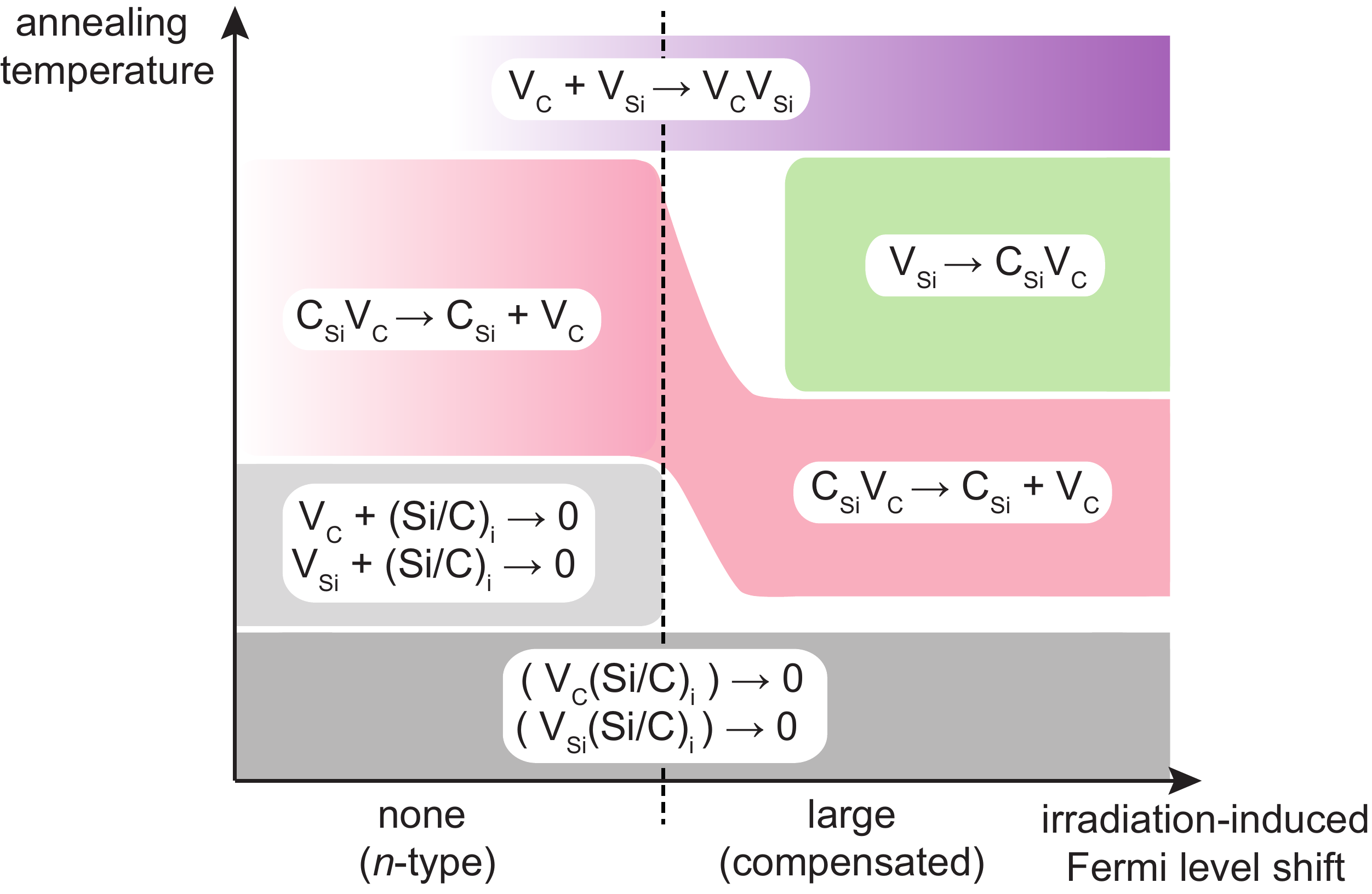}
	\caption{Proposed model for the interconversion of defect species created in irradiated $n$-type 4H-SiC. Color gradients encode concentration of the defects produced by the reactions (darker = higher conc.).}
	\label{fig:illustration}
\end{figure}

The model we propose for the annealing-induced conversion of intrinsic defects in irradiated $n$-type 4H-SiC is illustrated in Fig.~\ref{fig:illustration} and will now be summarized. At annealing temperatures of \SI{300}{\degreeCelsius} and below, closely spaced pairs of Si and C vacancies and interstitials (V\lwr{C}(Si/C)\lwr{i} and V\lwr{Si}(Si/C)\lwr{i}) annihilate easily in both $n$-type and compensated material. Increasing the temperature leads to continued recombination of non-nearest neighbor interstitial-vacancy pairs. This can only occur in $n$-type material because interstitial diffusion is severely hampered with increasingly positive charge state. The next annealing stage is the dissociation of carbon antisite-vacancy pairs via formation of individual C\lwr{Si} and V\lwr{C} defects. This process seems to be triggered more abruptly and at lower temperatures for compensated as compared to $n$-type samples. At even higher annealing temperatures, conversion of V\lwr{Si} to CAV pair occurs in the compensated samples. This process is visible because a large fraction of originally formed vacancies remained unannealed in the previous stage of blocked vacancy-interstitial recombination. For the highest temperatures in this experiment, the formation of divacancies via the diffusion of V\lwr{Si} and binding to V\lwr{C} defects is observed.

\section{Conclusion}

The defect population in proton-irradiated, $n$-type 4H-SiC epilayers and its evolution during post-annealing have been studied through a combination of optical and electrical methods. We have shown that the dominant defect species are silicon and carbon vacancies and interstitials, the carbon antisite-carbon vacancy pair (CAV), and the divacancy (VV). All these are already present after irradiation and dynamic annealing. Irradiation is also shown to lead to compensation of donors with increasing proton fluence, which can be partially reversed upon high-temperature annealing, supporting the notion of compensation by acceptor defects rather than by donor passivation. The presented data further support the assignment of the high-temperature EH\lwr{4} and EH\lwr{5} levels observed by deep level transient spectroscopy to the (+/0) CTL of the CAV pair in its four energetically inequivalent configurations. We show that a temperature-induced interconversion between the single Si vacancy (V\lwr{Si}) and the CAV pair, as it has been reported in $p$-type material, is suppressed under $n$-type conditions. Instead, CAV defects appear to anneal out via a different route that, at the same time, leads to an equally large increase of carbon vacancy (V\lwr{C}) concentration. We present a model describing this as a simple dissociation into isolated V\lwr{C} and carbon antisite (C\lwr{Si}) defects -- a process which, according to our data and model, could be activated at a much lower temperature than previously expected. It also seems to occur more abruptly and at lower temperatures when the Fermi level is lower in the band gap (under compensation). We furthermore show that until \SI{400}{\degreeCelsius}, the annealing behavior of irradiated $n$-type SiC is dominated by continued recombination of vacancies V\lwr{C}, V\lwr{Si} with self-interstitials, a process which is blocked in compensated SiC. Annealing at temperatures at \SI{800}{\degreeCelsius} and above lead to the diffusion-mediated formation of divacancies V\lwr{Si}V\lwr{C}.

\begin{acknowledgments}
	The authors are obliged to C. Zimmermann for supplying the software for simulating DLTS spectra. Financial support was kindly provided by the Research Council of Norway and the University of Oslo through the frontier research project FUNDAMeNT (no. 251131, FriPro ToppForsk-program). The Research Council of Norway is acknowledged for the support to the Norwegian Micro- and Nano-Fabrication Facility, NorFab, project number 245963. 
\end{acknowledgments}

\section*{Conflicts of interest}
There are no conflicts to declare.

\section*{References}

\bibliography{ref}

\end{document}